\documentclass[10pt,conference]{IEEEtran}
\IEEEoverridecommandlockouts
\usepackage{cite}
\usepackage{amsmath,amssymb,amsfonts}
\usepackage{algorithmic}
\usepackage{graphicx}
\usepackage{textcomp}
\usepackage{xcolor}
\usepackage[hyphens]{url}
\usepackage{fancyhdr}
\usepackage{hyperref}
\def\BibTeX{{\rm B\kern-.05em{\sc i\kern-.025em b}\kern-.08em
    T\kern-.1667em\lower.7ex\hbox{E}\kern-.125emX}}
\usepackage{xspace}
\usepackage{enumitem}
\usepackage{booktabs}
\usepackage{multirow}

\newcommand{\papername}{ADOR\xspace}
\makeatletter
\newcommand{\linebreakand}{%
  \end{@IEEEauthorhalign}
  \hfill\mbox{}\par
  \mbox{}\hfill\begin{@IEEEauthorhalign}
}
\makeatother

\begin{document}

\title{\papername: A Design Exploration Framework for LLM Serving with Enhanced Latency and Throughput\\
% {\footnotesize \textsuperscript{*}Note: Sub-titles are not captured in Xplore and should not be used}
% \thanks{Identify applicable funding agency here. If none, delete this.}
}

\author{
\IEEEauthorblockN{Junsoo Kim}
\IEEEauthorblockA{\textit{HyperAccel Inc.} \\
Seoul, South Korea \\
js.kim@hyperaccel.ai}
\and
\IEEEauthorblockN{Hunjong Lee}
\IEEEauthorblockA{\textit{HyperAccel Inc.} \\
Seoul, South Korea \\
hj.lee@hyperaccel.ai}
\and
\IEEEauthorblockN{Geonwoo Ko}
\IEEEauthorblockA{\textit{KAIST} \\
Daejeon, South Korea \\
geonwooko@kaist.ac.kr}
\linebreakand
\IEEEauthorblockN{Gyubin Choi}
\IEEEauthorblockA{\textit{HyperAccel Inc.} \\
Seoul, South Korea \\
gb.choi@hyperaccel.ai}
\and
\IEEEauthorblockN{Seri Ham}
\IEEEauthorblockA{\textit{KAIST} \\
Daejeon, South Korea \\
seri1215@kaist.ac.kr}
\and
\IEEEauthorblockN{Seongmin Hong}
\IEEEauthorblockA{\textit{HyperAccel Inc.} \\
Seoul, South Korea \\
sm.hong@hyperaccel.ai}
\and
\IEEEauthorblockN{Joo-Young Kim}
\IEEEauthorblockA{\textit{HyperAccel Inc.} \\
Seoul, South Korea \\
jy.kim@hyperaccel.ai}
}

\maketitle
\thispagestyle{plain}
\pagestyle{plain}

%%%%%%%%%%%%%%%%%%%%%%%%%%%%%%%%%%%%%%%%%%%%%%%%%%%%%%%%%%%%%%%%%%%%%%%%%%
%% Copyright HyperAccel, Inc. All Rights Reserved.
%% Section: Abstract
%%%%%%%%%%%%%%%%%%%%%%%%%%%%%%%%%%%%%%%%%%%%%%%%%%%%%%%%%%%%%%%%%%%%%%%%%%

\begin{abstract}

The growing adoption of Large Language Models (LLMs) across various domains has driven the demand for efficient and scalable AI-serving solutions. Deploying LLMs requires optimizations to manage their significant computational and data demands. The \textit{prefill} stage processes large numbers of input tokens in parallel, increasing computational load, while the \textit{decoding} stage relies heavily on memory bandwidth due to the auto-regressive nature of LLMs. Current hardware, such as GPUs, often fails to balance these demands, leading to inefficient utilization. While batching improves hardware efficiency, it delays response times, degrading Quality-of-Service (QoS).

This disconnect between vendors, who aim to maximize resource efficiency, and users, who prioritize low latency, highlights the need for a better solution. To address this, we propose \papername, a framework that automatically identifies and recommends hardware architectures tailored to LLM serving. By leveraging predefined architecture templates specialized for heterogeneous dataflows, \papername optimally balances throughput and latency. It efficiently explores design spaces to suggest architectures that meet the requirements of both vendors and users. \papername demonstrates substantial performance improvements, achieving 2.51$\times$ higher QoS and 4.01$\times$ better area efficiency compared to the A100 at high batch sizes, making it a robust solution for scalable and cost-effective LLM serving.

\end{abstract}

%%%%%%%%%%%%%%%%%%%%%%%%%%%%%%%%%%%%%%%%%%%%%%%%%%%%%%%%%%%%%%%%%%%%%%%%%%

%%%%%%%%%%%%%%%%%%%%%%%%%%%%%%%%%%%%%%%%%%%%%%%%%%%%%%%%%%%%%%%%%%%%%%%%%%
%% Copyright HyperAccel, Inc. All Rights Reserved.
%% Section: Introduction
%%%%%%%%%%%%%%%%%%%%%%%%%%%%%%%%%%%%%%%%%%%%%%%%%%%%%%%%%%%%%%%%%%%%%%%%%%

\section{Introduction}
\label{sec:intro}

% Importance of LLM Serving
Large Language Models (LLMs) have ushered in a paradigm shift across various industries, fundamentally transforming business operations and customer interactions. Originally designed for text generation, LLMs have rapidly evolved to support multimodal capabilities~\cite{yang2023dawn, liu2024visual, shao2023prompting}, including the ability to process and generate images, thereby extending their impact into previously specialized domains.
These technologies have been instrumental in automating complex tasks, generating high-quality output, and providing personalized user experiences at scale. As a result, the demand for efficient and scalable LLM serving solutions has grown significantly, necessitating innovative strategies to address the computational complexities and resource demands of these models. The continuous integration of LLMs into diverse sectors highlights their potential to further revolutionize industries, underscoring the need for robust serving architectures capable of meeting future demands.

%%%%%%%%%%%%%%%%%%%%%%%%%%%%%%%%%%%%%%%%%%%%%%%%%%%%%%%%%%%%%%%%%%%%%%%%%%
%% Figure 1 : Challenges at LLM Serving

\begin{figure}[t]
\centering
\includegraphics[width=1.0\linewidth]{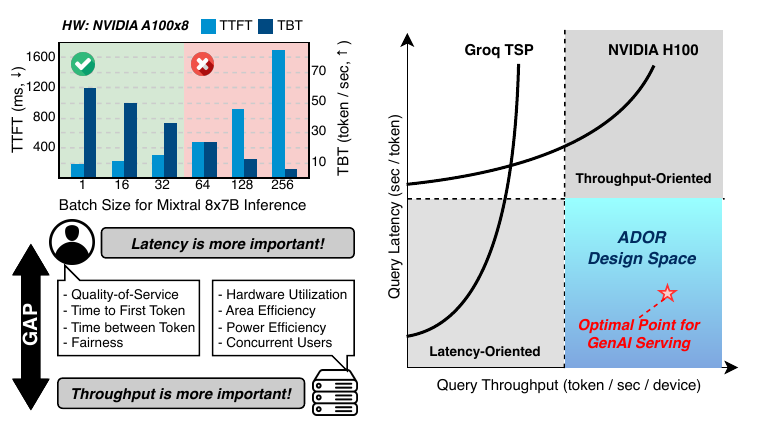}
\vspace{-0.25in}
\caption{Gap between end-user and vendor needs in LLM Serving. \papername explores and proposes hardware architectures that consider both throughput and latency.}
\label{fig:introduction}
\vspace{-0.20in}
\end{figure}

%%%%%%%%%%%%%%%%%%%%%%%%%%%%%%%%%%%%%%%%%%%%%%%%%%%%%%%%%%%%%%%%%%%%%%%%%%

% Challenges for LLM Serving
Despite the significant potential of LLM-based systems, several challenges persist in their deployment. Most modern LLMs adopt the Transformer~\cite{vaswani2017attention} architecture, inheriting its computational characteristics. The inference process in LLM can be broadly divided into two stages: the \textit{prefill} stage and the \textit{decoding} stage. The \textit{prefill} stage generates key-value pairs for input tokens, which can be processed in parallel between tokens. This parallelism allows for efficient computation, but as the number of tokens increases, the computational load scales significantly. In contrast, the \textit{decoding} stage involves generating tokens sequentially in an auto-regressive manner, where the model must repeatedly load data for each token generation~\cite{kwon2023efficient}. This stage is memory-bound, often leading to underutilization of compute units. To address these inefficiencies, batching techniques~\cite{kwon2023efficient} have been employed to enhance utilization rates. However, as shown in Fig.~\ref{fig:introduction}, increasing batch sizes can adversely affect Quality of Service (QoS). The architecture of Transformers restricts the sharing of key-value pairs across requests, leading to underutilization during attention block operations. Consequently, this limitation degrades critical QoS metrics, such as Time-to-First-Token (TTFT) and Time-Between-Tokens (TBT), ultimately constraining the effectiveness of batching.

As vendors aim to process more requests with fewer hardware resources, while end-users demand minimal batching to maintain high service quality, there is a significant gap between these two perspectives. Currently, there is a lack of hardware solutions that effectively bridge this gap. Vendors require hardware that maximizes throughput within a given budget or area, whereas end-users prioritize low latency for swift responses. The most representative hardware, GPUs~\cite{choquette2020nvidia}, offers lower throughput and latency efficiency due to their focus on programmability. Although GPUs utilize high-bandwidth memory (HBM)~\cite{jun2017hbm}, their bandwidth utilization hovers around 60\%, resulting in minimal performance improvement. Therefore, they are suboptimal for latency-sensitive workloads. On the other hand, specialized hardware like Google’s TPU~\cite{jouppi2023tpu} and Groq’s TSP~\cite{abts2020think, abts2022groq} are optimized for throughput and latency, respectively, but neither adequately addresses both requirements, failing to bridge the gap between the vendor's needs and the end user.

% Propose this paper
As illustrated in Fig.~\ref{fig:introduction}, serving LLMs requires an architecture that effectively balances both throughput and latency within the given hardware specifications. In this paper, we propose the Automatic Dataflow Optimization and Exploration (\textbf{\papername}) framework, which is designed to identify and explore optimal hardware architectures for efficiently serving LLMs, thereby bridging the gap between vendors and end-users. The \papername framework offers the following key features:

\begin{description}[labelindent=0.0em,nolistsep,leftmargin=1.5em]
\item[(1)] \textbf{Optimal Architecture Identification:}
\papername identifies an architecture that can effectively handle both throughput and latency within given hardware specifications such as area, memory bandwidth, and P2P bandwidth. It is based on a Heterogeneous Dataflow Architecture (HDA)\cite{yang2020co, kwon2021heterogeneous}, which optimally balances the proportions of throughput-oriented systolic arrays, latency-oriented multiplier-accumulator (MAC) trees, and versatile vector units. This balanced HDA design ensures efficient management of both latency and throughput.
\item[(2)] \textbf{Latency and Throughput Optimization through Unified Dataflows:}
\papername achieves optimization by carefully selecting and integrating dataflows tailored for both latency-oriented and throughput-oriented workloads. For latency-oriented operations, \papername leverages lightweight, low-latency dataflows that minimize memory access overhead and maximize computation-communication overlap. For throughput-oriented operations, it employs highly parallel dataflows optimized for efficient utilization of large-scale systolic arrays. By unifying these distinct dataflows into a cohesive framework, \papername enables seamless transitions between latency-critical and throughput-intensive scenarios, ensuring benefits in both cases. This unified dataflow approach not only balances the hardware’s operational efficiency but also avoids additional hardware complexity, making the solution both effective and scalable.
\item[(3)] \textbf{Efficient Serving Scheduling Methods:}
\papername{} offers scheduling methods to ensure the efficient operation of the HDA structure in real serving environments. The scheduling handles \textit{prefill} and \textit{decoding} stages simultaneously and proposes efficient serving methods for various GenAI models. It also simulates real-world serving environments to predict the expected hardware utilization and QoS for the proposed hardware architecture.
\end{description}

\noindent
In summary, \papername aims to bridge the gap between vendors and end-users by proposing a comprehensive framework that optimizes hardware architectures for efficient LLM serving, addressing both throughput and latency challenges.

%%%%%%%%%%%%%%%%%%%%%%%%%%%%%%%%%%%%%%%%%%%%%%%%%%%%%%%%%%%%%%%%%%%%%%%%%%
%%%%%%%%%%%%%%%%%%%%%%%%%%%%%%%%%%%%%%%%%%%%%%%%%%%%%%%%%%%%%%%%%%%%%%%%%%
%% Copyright HyperAccel, Inc. All Rights Reserved.
%% Section: Backgrounds
%%%%%%%%%%%%%%%%%%%%%%%%%%%%%%%%%%%%%%%%%%%%%%%%%%%%%%%%%%%%%%%%%%%%%%%%%%

\section{Background}
\label{sec:background}

%%%%%%%%%%%%%%%%%%%%%%%%%%%%%%%%%%%%%%%%%%%%%%%%%%%%%%%%%%%%%%%%%%%%%%%%%%
%% Figure 2 : Overview of LLM Serving

\begin{figure*}[t]
\centering
\includegraphics[width=1.0\linewidth]{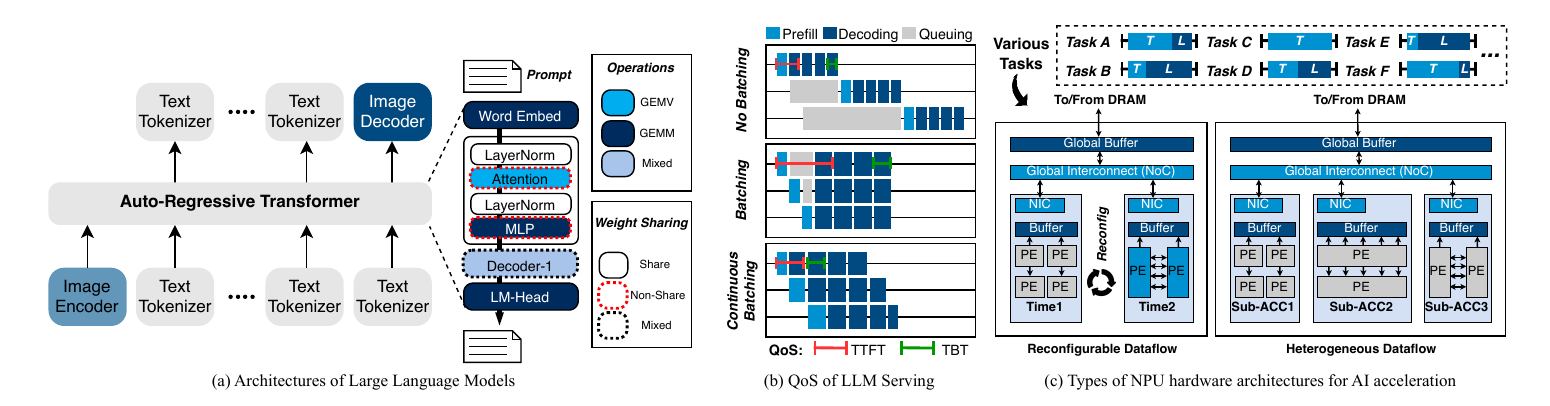}
\vspace{-0.25in}
\caption{(a) Architecture of Large Language Models. (b) Variations in TTFT and TBT types based on batching methods. (c) The structure and differences between Coarse-Grained Reconfigurable Architecture (CGRA) and Heterogeneous Dataflow Architecture (HDA).}
\label{fig:background}
\vspace{-0.20in}
\end{figure*}

%%%%%%%%%%%%%%%%%%%%%%%%%%%%%%%%%%%%%%%%%%%%%%%%%%%%%%%%%%%%%%%%%%%%%%%%%%

% Large Language Models
\subsection{Overview of Large Language Models}
\label{sec:background:overview}

LLMs, based on the Transformer architecture~\cite{vaswani2017attention} as shown in Fig.\ref{fig:background}-(a), include models like OpenAI’s GPT series~\cite{brown2020language}, Google’s Gemini~\cite{team2023gemini}, and Meta’s LLaMA~\cite{touvron2023llama}, designed for human-like text generation.
LLM inference consists of \textbf{\textit{prefill}} and \textbf{\textit{decoding}} stages~\cite{zhou2024survey}. The \textit{prefill} stage generates key-value pairs for input tokens, using parallel GEMM operations that scale with token length, increasing compute demands. In the \textit{decoding} stage, tokens are generated sequentially, requiring repeated GEMV operations to load model parameters and key-value pairs. This stage is memory bandwidth-intensive due to the large size of models and key-value pairs.

%%%%%%%%%%%%%%%%%%%%%%%%%%%%%%%%%%%%%%%%%%%%%%%%%%%%%%%%%%%%%%%%%%%%%%%%%%
%% Figure 3 : Key-Value Caching

\begin{figure}[t]
\centering
\includegraphics[width=1.0\linewidth]{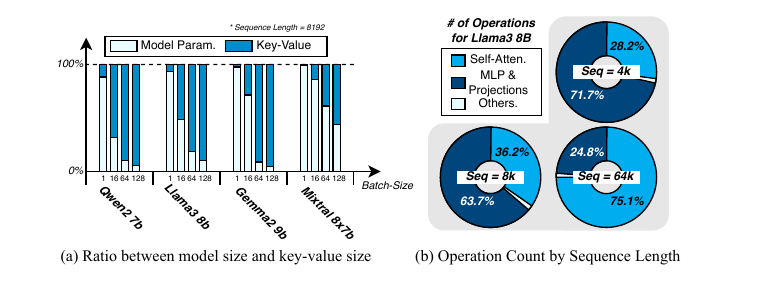}
\vspace{-0.25in}
\caption{(a) Proportion of key-value cache size for various models. As batch size increases, the proportion of key-value size grows. (b) Proportion of Attention operations in various LLM models.}
\label{fig:key_value_cache}
\vspace{-0.20in}
\end{figure}

%%%%%%%%%%%%%%%%%%%%%%%%%%%%%%%%%%%%%%%%%%%%%%%%%%%%%%%%%%%%%%%%%%%%%%%%%%

% Characteristics for LLM
\subsection{Characteristics of LLM Serving}
\label{sec:background:serving}

% Batching in Generative AI Serving
Typical GPUs~\cite{choquette2020nvidia, choquette2022nvidia} or NPUs~\cite{jouppi2023tpu} are designed with a high number of compute units, making them relatively effective at handling the \textit{prefill} stage. However, their memory bandwidth does not match this high computational capacity. As a result, during the \textit{decoding} stage, these devices often encounter hardware underutilization due to this imbalance~\cite{chitty2024llm}. To address this issue, batching is employed in real-world Generative AI serving. Batching enables parallelism in the \textit{decoding} stage along the batch dimension, thus mitigating hardware underutilization to some extent. Additionally, techniques such as continuous batching\cite{yu2022orca} can be utilized to process the \textit{decoding} stage of one request simultaneously with the \textit{prefill} stage of the next request, thereby maintaining high utilization levels.

% Key-Value Pair
However, not all parameters can be shared during batching. LLM models use key-value pairs that are generated during runtime to perform inference. Since these key-value pairs are unique to each request, they cannot be shared during batching. Consequently, when the attention block is executed, it demands high memory bandwidth again, leading to hardware underutilization. As shown in Fig.~\ref{fig:key_value_cache}-(a), in recent models with a batch size of 128, over 90\% of the data that needs to be read from DRAM pertains to key-value pairs. This significantly impacts the latency for generating each token. When the batch size increases, the speed of processing attention during the \textit{decoding} stage slows down~\cite{xiao2023efficient}, negatively impacting the TBT metric. If continuous batching is additionally applied, it further affects TTFT, as the \textit{prefill} and \textit{decoding} stages occur simultaneously. This combination significantly degrades the QoS of LLM serving. As shown in Fig.\ref{fig:key_value_cache}-(b), the proportion of Attention operations accounts for around 25\% in several models, indicating that these issues are prevalent across all LLM models.

% Heterogeneous Dataflow Architecture
\subsection{DNN Hardware Architectures for Different Workloads}
\label{sec:background:hda}

LLM serving demands a hardware architecture that can efficiently handle the conflicting requirements of \textit{prefill} and \textit{decoding}. To address similar issues, two approaches have recently been proposed: Coarse-Grained Reconfigurable Architecture (CGRA) and Heterogeneous Dataflow Architecture (HDA). CGRA is designed so that a single core can handle both modes effectively by changing its configuration at runtime through control signals. This method allows a single core to handle various tasks with high utilization regardless of the workload. However, to support multiple modes, CGRA requires additional logic such as switches and wires, making the control complex and resulting in less area efficiency and poorer power efficiency.
To overcome these challenges, recent research has shown that using HDA~\cite{kwon2021heterogeneous} can effectively address issues in multi-DNN scenarios. HDA arranges multiple cores with different dataflows to handle various tasks through workload scheduling. Each core in an HDA does not require additional logic for control, maintaining area and power efficiency. The paper reported that under the same conditions, HDA achieved up to 80.4\% latency improvement and 41.3\% power consumption savings compared to CGRA. This indicates that HDA is more effective than CGRA for efficiently handling two different types of workloads.

%%%%%%%%%%%%%%%%%%%%%%%%%%%%%%%%%%%%%%%%%%%%%%%%%%%%%%%%%%%%%%%%%%%%%%%%%%
%%%%%%%%%%%%%%%%%%%%%%%%%%%%%%%%%%%%%%%%%%%%%%%%%%%%%%%%%%%%%%%%%%%%%%%%%%
%% Copyright HyperAccel, Inc. All Rights Reserved.
%% Section: Motivation
%%%%%%%%%%%%%%%%%%%%%%%%%%%%%%%%%%%%%%%%%%%%%%%%%%%%%%%%%%%%%%%%%%%%%%%%%%

\section{Motivation}
\label{sec:motivation}

% Overview of this Section
The goal of this paper is to present a novel framework for exploring and proposing a hardware architecture that bridges the gap between vendors and end-users in LLM serving. As previously mentioned in Section~\ref{sec:background:serving}, LLM demands both high throughput and low latency. In this section, we will analyze the characteristics of the hardware currently used for LLM serving and, based on this analysis, discuss the direction for finding efficient hardware architecture.

%%%%%%%%%%%%%%%%%%%%%%%%%%%%%%%%%%%%%%%%%%%%%%%%%%%%%%%%%%%%%%%%%%%%%%%%%%
%% Table 1 : Analysis of Current Serving Hardware

\begin{table}[t]
    \centering
    \caption{Analysis of Current Serving Hardware}
    \begin{tabular}{lccc}
        \toprule
        \multirow{2}{*}{\textbf{Key Specifications}} & \textbf{NVIDIA} & \textbf{Google} & \textbf{Groq} \\
        & \textbf{H100} & \textbf{TPUv4} & \textbf{TSP} \\
        \midrule
        Core Frequency (MHz) & 1593 & 1050 & 1000 \\
        Technology & 4nm & 7nm & 14nm \\
        Peak Performance (TFLOPS) & 1000 & 275 & 205 \\
        On-chip SRAM (MB) & 80 & 160 & 220 \\
        DRAM Type & HBM3e & HBM2 & (SRAM) \\
        DRAM Size (GB) & 80 & 32 & (0.22) \\
        Memory Bandwidth (GB/s) & 3350 & 1200 & (80000) \\
        P2P Bandwidth (GB/s) & 900 & 200 & 330 \\
        Maximum TDP (W) & 700 & 275 & 300 \\
        Die Size (mm$^2$) & 814 & 400 & 725 \\
        \bottomrule
    \end{tabular}
    \label{tab:compare_uarch}
    \vspace{-0.20in}
\end{table}

%%%%%%%%%%%%%%%%%%%%%%%%%%%%%%%%%%%%%%%%%%%%%%%%%%%%%%%%%%%%%%%%%%%%%%%%%%

% Analysis Current Hardware
\subsection{Limitations of Current Serving Hardware}
\label{sec:motivation:analysis}

\begin{figure}[t]
\centering
\includegraphics[width=1.0\linewidth]{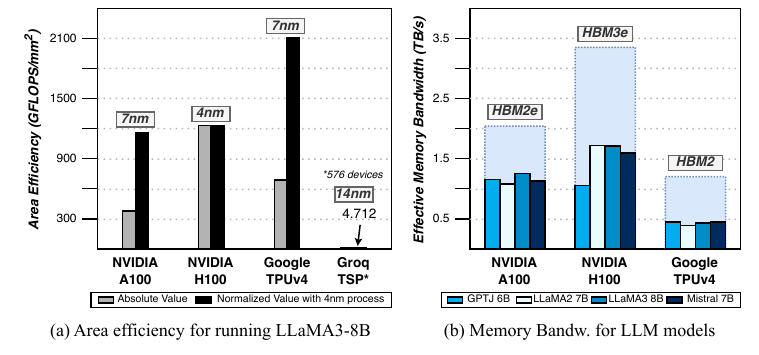}
\vspace{-0.25in}
\caption{(a) Average computational performance per unit area for various hardware during the \textit{prefill} stage of LLaMA3 8B. (b) Actual memory bandwidth utilization for various GenAI models. Both GPU and TPU show less than 60\% utilization compared to their specifications.}
\label{fig:limitations_hardware}
\vspace{-0.20in}
\end{figure}

% Analysis & Limitations of NVIDIA GPU
NVIDIA GPUs~\cite{choquette2020nvidia, choquette2022nvidia} are currently the primary hardware used for serving LLM. GPUs boast high computational power and utilize HBM to deliver data with high memory bandwidth. On the surface, the hardware specifications suggest that GPUs should meet to both throughput and latency needs.
However, the Simultaneous Multi-Threading (SMT) architecture~\cite{tullsen1995simultaneous} inherent in GPUs is not sufficient to adequately address both throughput and latency requirements. From a latency perspective, weights should be loaded into the computational cores as quickly as possible, processed, and then immediately replaced by the next set of weights, especially for attention blocks. Although HBM provides fast data transfer from memory, the data does not reach the CUDA cores efficiently due to the complex control processes inherent in SMT. As a result, GPUs exhibit low memory bandwidth utilization, achieving less than 60\% efficiency during the \textit{decoding} stage, as shown in Fig.~\ref{fig:limitations_hardware}-(b).

On the other hand, from a throughput perspective, GPUs are not purely throughput-oriented. Unlike architectures such as systolic arrays, GPUs require control logic for each core. This requirement exists because each core must be able to handle threads independently. As a result, GPUs are forced to sacrifice both throughput-oriented and latency-oriented features in order to achieve high programmability. While programmability is essential for training models and accelerating various applications beyond LLM, it becomes less of an advantage for LLM serving. In the context of LLM serving, throughput-oriented and latency-oriented features are more critical. Therefore, despite their versatility, GPUs are not ideally suited for LLM serving.

%%%%%%%%%%%%%%%%%%%%%%%%%%%%%%%%%%%%%%%%%%%%%%%%%%%%%%%%%%%%%%%%%%%%%%%%%%
%% Figure 4 : Limitations of Current Serving Hardware

%%%%%%%%%%%%%%%%%%%%%%%%%%%%%%%%%%%%%%%%%%%%%%%%%%%%%%%%%%%%%%%%%%%%%%%%%%

% Analysis & Limitations of NPU (Google TPU and Groq TSP)
Recently, there has been a move towards using NPUs for LLM serving. However, current NPUs also fail to fully meet both throughput and latency requirements. Google’s TPU, which adopts a systolic array architecture, integrates a high number of compute units per area, making it throughput-oriented. Nevertheless, due to the inherent characteristics of systolic arrays they exhibit low efficiency for GEMV operations where latency is critical. In contrast, Groq’s TSP loads all weights into on-chip SRAM, providing 80 TB/s of memory bandwidth in each chip, and uses a streaming flow to rapidly supply data to the compute units. However, TSP’s requirement to store tens of gigabytes of model parameters on-chip necessitates the use of a large number of chips. From a throughput perspective, this results in an overabundance of compute units, and achieving 100\% efficiency with such a large number of units is extremely challenging. Additionally, from a vendor’s perspective, it is not economically viable to use a significant number of devices to serve a single model. As shown in Fig.~\ref{fig:limitations_hardware}, although TPU has higher area efficiency than GPU in serving the LLaMA3 8B, its memory bandwidth utilization is worse compared to the GPU. On the other hand, TSP’s use of numerous chips results in low area efficiency.

% Considerations for Serving Hardware
\subsection{Architectural Considerations for Serving Hardware}
\label{sec:motivation:consider}

\begin{figure}[t]
\centering
\includegraphics[width=1.0\linewidth]{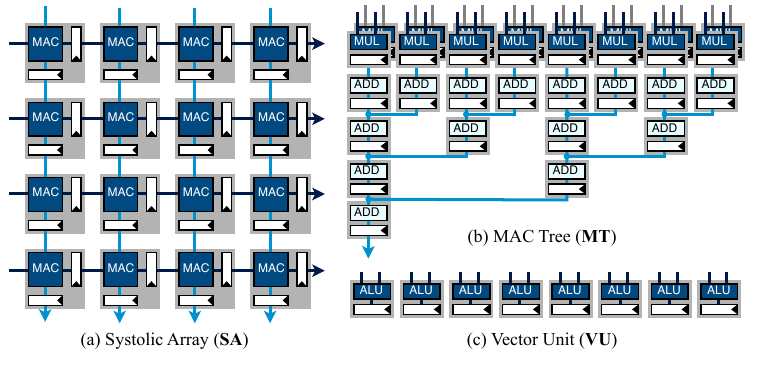}
\vspace{-0.25in}
\caption{Architectural and dataflow differences of three micro-architectures: (a) systolic array, (b) MAC tree, and (c) vector unit.}
\label{fig:compare_uarch}
\vspace{-0.10in}
\end{figure}

As discussed in Section~\ref{sec:background:hda}, HDA presents a promising solution to the limitations of current hardware for LLM. To understand the effective implementation of HDA, we will delve into the hardware architectures prevalent in modern NPUs and explore their respective strengths and weaknesses. As illustrated in Fig.~\ref{fig:compare_uarch}, modern NPUs utilize two major hardware architectures, systolic array (\textbf{SA}) and MAC tree (\textbf{MT}).

\textbf{Systolic array} is widely used in various NPUs. SA is specialized for GEMM operations, maximizing data locality. As summarized in Table~\ref{tab:compare_uarch}, SA is highly throughput-oriented and provides excellent area and energy efficiency. However, for GEMV operations, SA appears to be suboptimal~\cite{samajdar2018scale, heo2024neupims}. As the size of the SA increases, the latency also increases due to the diagonal distribution of input data, resulting in lower utilization of compute units. Additionally, weight double buffering is not feasible in this case, exposing pre-fetch latency and making SA less suitable for latency-sensitive workloads.

\textbf{MAC tree} is optimized for dot-product operations. They multiply each element of the vectors and accumulate the results using the adder tree. Since GEMM and GEMV operations consist of multiple dot-products, they can also be handled by MTs. MTs can process fetched inputs and weights immediately, providing low-latency results. As compared in Table~\ref{tab:compare_uarch}, MTs are highly latency-oriented, offering low latency for the same compute units. However, due to the tree-based structure, MTs have lower compute unit density in the physical implementation compared to SAs. This results in lower computational density and economic inefficiency in terms of throughput, making MTs less suitable for throughput-sensitive workloads.

As discussed above, effectively serving LLM requires leveraging these hardware architectures appropriately. To efficiently determine the optimal ratio of these components for given workloads, this paper outlines the overall template architecture and design space in Section~\ref{sec:design_space} and explains how to find the optimal architecture in Section~\ref{sec:arch_search}. Additionally, in Section~\ref{sec:evaluation}, we execute the \papername framework to evaluate and analyze the performance of the proposed hardware architectures.

%%%%%%%%%%%%%%%%%%%%%%%%%%%%%%%%%%%%%%%%%%%%%%%%%%%%%%%%%%%%%%%%%%%%%%%%%%
%% Figure 5 : Compare systolic array vs mac tree vs vector unit

%%%%%%%%%%%%%%%%%%%%%%%%%%%%%%%%%%%%%%%%%%%%%%%%%%%%%%%%%%%%%%%%%%%%%%%%%%
%% Table 2 : Compare systolic array vs mac tree

\begin{table}[t]
    \centering
    \caption{Key features of Systolic array and MAC tree}
    \begin{tabular}{lccc}
        \toprule
        & \textbf{Systolic Array} & \textbf{MAC Tree}\\
        \midrule
        Target Operation & Matrix-Multiplication & Dot-product \\
        Compute Intensity & High & Relatively-Low \\
        Overall Latency & Relatively-High & Low \\
        Suitable Workload & Throughput-sensitive & Latency-sensitive \\
        \bottomrule
    \end{tabular}
    \label{tab:compare_uarch}
    \vspace{-0.20in}
\end{table}

%%%%%%%%%%%%%%%%%%%%%%%%%%%%%%%%%%%%%%%%%%%%%%%%%%%%%%%%%%%%%%%%%%%%%%%%%%
%%%%%%%%%%%%%%%%%%%%%%%%%%%%%%%%%%%%%%%%%%%%%%%%%%%%%%%%%%%%%%%%%%%%%%%%%%
%% Copyright HyperAccel, Inc. All Rights Reserved.
%% Section: Design Space
%%%%%%%%%%%%%%%%%%%%%%%%%%%%%%%%%%%%%%%%%%%%%%%%%%%%%%%%%%%%%%%%%%%%%%%%%%

\section{\papername Design Space}
\label{sec:design_space}

%%%%%%%%%%%%%%%%%%%%%%%%%%%%%%%%%%%%%%%%%%%%%%%%%%%%%%%%%%%%%%%%%%%%%%%%%%
%% Figure 7 : ADOR's Design Space

\begin{figure*}[t]
\centering
\includegraphics[width=1.0\linewidth]{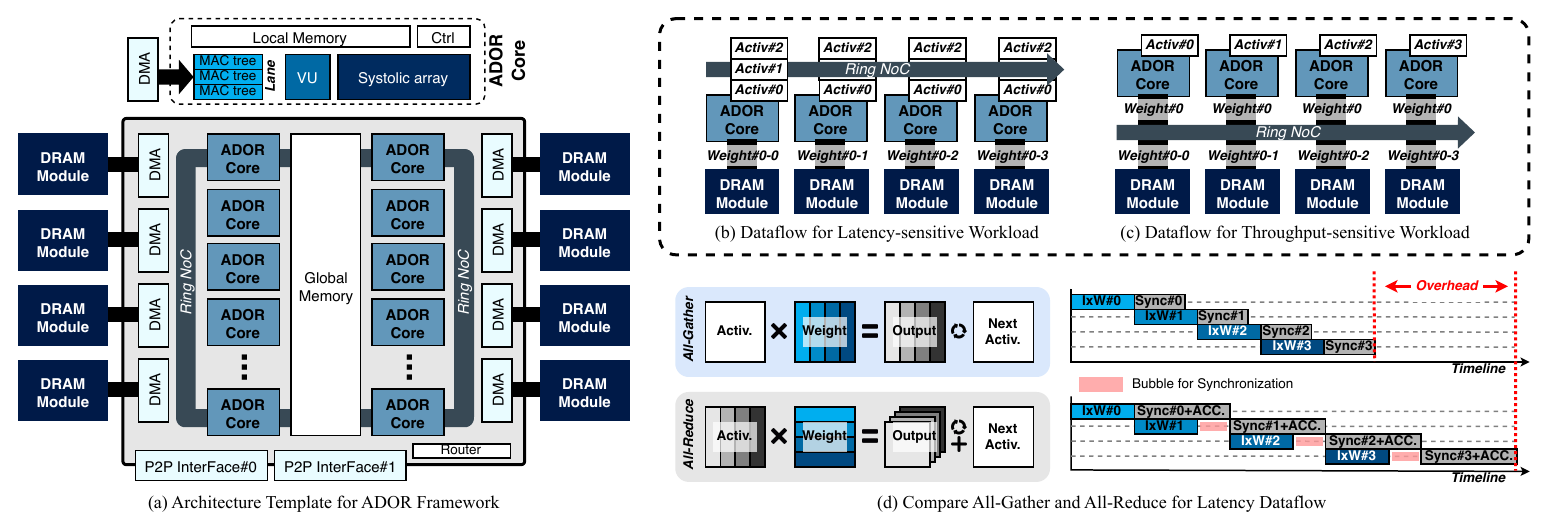}
\vspace{-0.25in}
\caption{(a) \papername framework’s architecture template considering both throughput and latency-sensitive workloads. (b) Multi-core dataflow for latency-sensitive workloads, where each core processes the same activation with different weights. (c) Multi-core dataflow for throughput-sensitive workloads, where each core processes different activations with the same weights. (d) Two methods for synchronization in latency-oriented dataflow. When computation and communication overlap, all-gather is more advantageous than all-reduce.}
\label{fig:design_space}
\vspace{-0.20in}
\end{figure*}

%%%%%%%%%%%%%%%%%%%%%%%%%%%%%%%%%%%%%%%%%%%%%%%%%%%%%%%%%%%%%%%%%%%%%%%%%%

As discussed in Section~\ref{sec:motivation}, serving LLMs effectively requires architectures that balance throughput and latency. \papername utilizes predefined templates, including systolic arrays, MAC trees, and associated dataflows, along with NoC and P2P configurations. Based on workloads from users and vendors, \papername determines hardware parameters to propose an optimal LLM architecture. This section defines the design space of \papername, encompassing its architecture templates.

% Order
% 1. Systolic array & MAC tree
% 2. Local memory & global memory
% 3. Multi-core dataflow
% 4. Mutli-device dataflow
% 5. Scheduling

% Design space for compute units
\subsection{Compute Units}
\label{sec:design_space:sa_mt}

\textbf{Latency-oriented Design.}
To begin, let’s examine the latency-oriented architecture. As previously discussed, latency depends on how efficiently the available memory bandwidth is utilized. In LLM computations, GEMV operations have a significant impact on latency~\cite{heo2024neupims}, especially as the key and value sizes grow and their share of the total computation increases. Therefore, it is important to quickly consume the key-value pairs fetched from DRAM in order to proceed to the next computation~\cite{prabhu2024vattention}. To achieve this, \papername employs a MAC tree architecture that allows weights read from DRAM to be fed directly into the compute units without first being storing in SRAM. This approach ensures that the data is promptly processed, minimizing latency~\cite{moon2023hyperaccel}.

\textbf{Throughput-oriented Design.}
Conversely, throughput-sensitive workloads require a different approach than latency-oriented scenarios. To maximize throughput, \papername primarily utilizes systolic arrays, similar to other NPUs. As discussed in Section~\ref{sec:motivation:consider}, systolic arrays are highly effective at increasing throughput due to their high compute unit density relative to area. Given the characteristics of LLM, where model parameters are typically very large, weights are stored in DRAM, while the relatively smaller activations are stored on-chip. Due to the inherent structure of systolic arrays, the speed at which data is fed impacts throughput. Therefore, \papername employs a Weight Stationary systolic array that pre-fetches weights and supplies activations to ensure high throughput.

% Local memory and Global memory
\subsection{Local memory \& Global memory}
\label{sec:design_space:on_chip_memory}

\textbf{Latency-oriented Design.}
\papername’s architecture template includes two types of on-chip memory: local memory and global memory. Local memory, which is allocated to each core, is responsible for storing activations. To ensure optimal performance, the bandwidth of the off-chip memory should only be used to fetch weights. Therefore, the local memory must be large enough to store all activations for a single layer. However, during the \textit{decoding} stage, local memory usage is generally lower compared to the \textit{prefill} stage~\cite{zhangllmcompass}, so this does not significantly affect performance. Global memory is used for sharing activations among multiple cores. A typical example is the attention mechanism in the \textit{prefill} stage. The key-value pairs generated in the current chunk can be stored in global memory instead of accessing DRAM, allowing some key-value pairs in attention to proceed without consuming DRAM bandwidth. This increases the effective memory bandwidth and reduces latency.

\textbf{Throughput-oriented Design.}
During GEMM operations, having more activations on-chip increases the reusability of weights, so it is ideal for the local memory to be as large as possible. However, since this is model-dependent, it is impossible to store all activations for large models~\cite{touvron2023llama} with large maximum token lengths~\cite{team2023gemini}. In such cases, activations can be tiled along the token (or row of the matrix) for computation. However, if activations for a single token cannot be stored, access to DRAM becomes necessary. Therefore, it is critical to configure the local memory to ensure that activations for at least one token can be stored based on the model’s information.

% Mutli-core dataflow
\subsection{Dataflow for Multi-Core}
\label{sec:design_space:multi-core}

\textbf{Latency-oriented Design.}
Achieving low latency requires optimizing the multi-core dataflow of \papername. One method for tiling GEMV operations involves each core processing the same weight matrix while partitioning the input matrix row-wise, as shown in Fig.~\ref{fig:design_space}-(c). Weights are broadcast from DRAM to all cores, eliminating the need for synchronization. However, broadcasting incurs latency penalties due to physical implementation constraints and requires the NoC to match HBM’s TB/s bandwidth, increasing design complexity and reducing area and power efficiency.

The second method involves each core holding the same input matrix while computing different parts of the weight matrix, requiring synchronization between cores for the output matrix before the next GEMV operation. Latency can be minimized if the compiler ensures each core fetches data from the nearest DRAM module, avoiding memory bandwidth loss and reducing physical design complexity. However, when performing computations with different weights across cores, it is necessary to synchronize the activation results for the next GEMV operation. For synchronization, two methods are commonly used: all-gather and all-reduce. As shown in Figure~\ref{fig:design_space}-(d), all-gather synchronizes small final-sum results after GEMV computation, allowing pipelining to overlap computation and communication, effectively hiding synchronization overhead. In contrast, all-reduce accumulates larger partial-sum results, requiring more data transfer and additional accumulation steps, leading to higher latency and increased NoC bandwidth demands. Due to these advantages, \papername adopts the all-gather method for core synchronization.

\textbf{Throughput-oriented Design.}
In latency-oriented designs, having each core hold the same activation is beneficial for minimizing latency. However, for multi-core computations, this approach necessitates that all cores store the same input activations, which can be inefficient. GEMM operations inherently require significant input activation storage, but local memory is often insufficient to hold all the necessary data.

While the previously discussed method is effective for reducing latency, it is not optimal for throughput-sensitive workloads, where latency is less critical. Therefore, using the method illustrated in Fig.~\ref{fig:design_space}-(b) is more suitable for these workloads. When pre-fetching weights in systolic arrays, the use of double buffering can effectively hide latency even when weights are fetched from physically distant DRAM modules. This allows broadcasting the same weights to all cores without idling compute units, ensuring full utilization.

%%%%%%%%%%%%%%%%%%%%%%%%%%%%%%%%%%%%%%%%%%%%%%%%%%%%%%%%%%%%%%%%%%%%%%%%%%
%% Figure 8 : Multi-Device

\begin{figure}[t]
\centering
\includegraphics[width=1.0\linewidth]{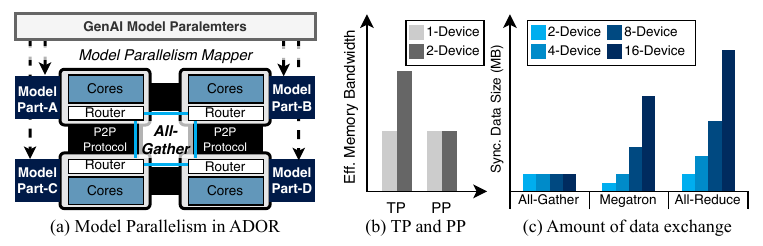}
\vspace{-0.25in}
\caption{(a) Methods of model parallelism in multi-device inference. (b) Differences between Tensor Parallelism (TP) and Pipeline Parallelism (PP) in terms of memory bandwidth. (c) Relationship between each TP method and the data exchange volume between devices.}
\label{fig:multi_device}
\vspace{-0.20in}
\end{figure}

%%%%%%%%%%%%%%%%%%%%%%%%%%%%%%%%%%%%%%%%%%%%%%%%%%%%%%%%%%%%%%%%%%%%%%%%%%

% Multi-Device Template
\subsection{Multi-Device Mapping for Large Models}
\label{sec:multi_device}

In LLM serving, limited memory capacity and bandwidth often necessitate multi-device computing through model parallelism. The two primary methods are Tensor Parallelism (TP) and Pipeline Parallelism (PP). TP splits the weight matrix of a GEMM (or GEMV) operation across devices, synchronizing with all-gather or all-reduce, while PP assigns entire layers to individual devices and passes results sequentially. Both methods utilize the memory and throughput of multiple devices, but they differ in latency. As shown in Fig.~\ref{fig:multi_device}-(b), TP reduces latency by distributing computations across devices, whereas PP provides no latency benefits due to pipelining. While TP introduces synchronization overhead, overlapping computation and communication mitigates this, making it more suitable for LLM serving where both throughput and latency are critical.

In TP, synchronization can be achieved through all-gather, all-reduce, or a hybrid Megatron~\cite{shoeybi2019megatron} approach. Megatron reduces synchronization steps by combining all-gather and all-reduce sequentially. However, all-reduce requires more data transfer, as it involves sending partial sums of the entire data. As shown in Fig.\ref{fig:multi_device}-(c), all-gather maintains a constant data volume up to 16 devices, whereas all-reduce scales with the number of devices. While Megatron is efficient with fewer devices, it demands higher bandwidth as device count increases, making all-gather more suitable for larger setups. Fig.\ref{fig:multi_device}-(a) illustrates how \papername identifies the minimum P2P bandwidth needed to overlap computation and communication effectively using all-gather.

%%%%%%%%%%%%%%%%%%%%%%%%%%%%%%%%%%%%%%%%%%%%%%%%%%%%%%%%%%%%%%%%%%%%%%%%%%
%% Figure 6 : ADOR's Scheduling Method

\begin{figure}[t]
\centering
\includegraphics[width=1.0\linewidth]{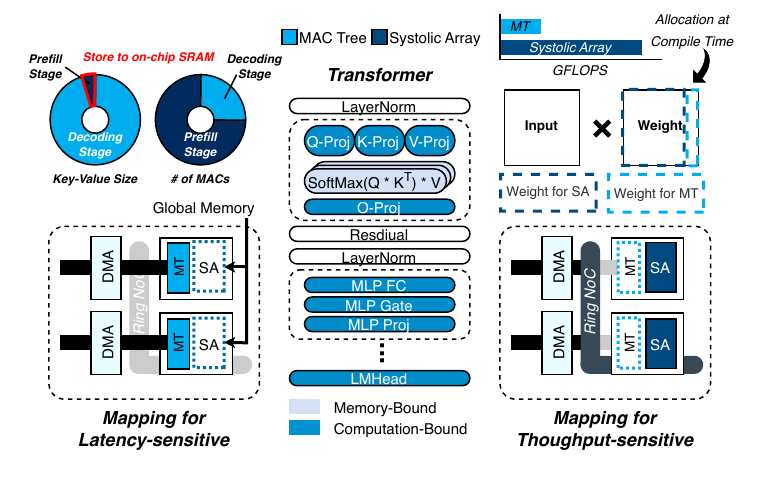}
\vspace{-0.25in}
\caption{Scheduling methods for \papername’s HDA. For latency-sensitive workloads, MAC tree uses all DRAM bandwidth for GEMV. For throughput-sensitive workloads, both the systolic array and MAC tree perform GEMM operations.}
\label{fig:scheduling}
\vspace{-0.20in}
\end{figure}

%%%%%%%%%%%%%%%%%%%%%%%%%%%%%%%%%%%%%%%%%%%%%%%%%%%%%%%%%%%%%%%%%%%%%%%%%%

% Dynamic Scheduling Method for HDA
\subsection{Dynamic Scheduling Method for LLM Serving}
\label{sec:dynamic_scheduling}

\textbf{Latency-oriented Design.}
As illustrated in Fig.~\ref{fig:scheduling}, a optimized scheduling method is necessary to efficiently utilize systolic arrays and MAC trees in actual LLM serving. For latency-sensitive workloads, maximizing memory bandwidth efficiency is crucial. Therefore, MAC trees are used exclusively to perform GEMV operations. During the \textit{decoding} stage, while the MAC tree is handling the attention with full use of the DRAM bandwidth, the systolic array utilizes key-value pairs stored in global memory, ensuring it does not interfere with DRAM bandwidth. If the key-value pairs are in DRAM, the systolic array waits until the \textit{decoding} stage is complete before proceeding.

\textbf{Throughput-oriented Design.}
When token lengths are sufficient or batch sizes are large enough, systolic arrays are employed to quickly handle the high computation load. Furthermore, since MAC trees can also perform GEMM operations, they can be used alongside systolic arrays to maximize the utilization of compute units within the chip. The design typically assigns fewer compute units to MAC trees than to systolic arrays because MAC trees are configured to handle only the data throughput from memory bandwidth. Therefore, considering the ratio of compute units between systolic arrays and MAC trees, the workload distribution for GEMM operations is determined at compile time and applied during scheduling to ensure efficient utilization of the available computational resources.

%%%%%%%%%%%%%%%%%%%%%%%%%%%%%%%%%%%%%%%%%%%%%%%%%%%%%%%%%%%%%%%%%%%%%%%%%%
%%%%%%%%%%%%%%%%%%%%%%%%%%%%%%%%%%%%%%%%%%%%%%%%%%%%%%%%%%%%%%%%%%%%%%%%%%
%% Copyright HyperAccel, Inc. All Rights Reserved.
%% Section: Architecture Searching Methodology
%%%%%%%%%%%%%%%%%%%%%%%%%%%%%%%%%%%%%%%%%%%%%%%%%%%%%%%%%%%%%%%%%%%%%%%%%%

\section{\papername Architecture Searching}
\label{sec:arch_search}

\begin{figure}[t]
\centering
\includegraphics[width=1.0\linewidth]{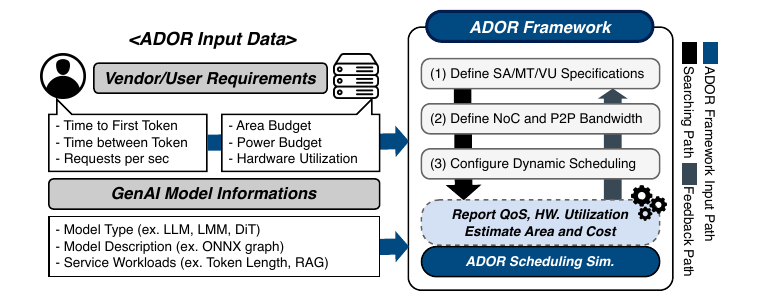}
\vspace{-0.25in}
\caption{ADOR framework's input parameters and architecture searching flow}
\label{fig:searching}
\vspace{-0.20in}
\end{figure}

This section explains how \papername searches for the optimal architecture to balance throughput and latency-sensitive workloads based on a given template architecture. The framework collects SLAs from end-users, and hardware constraints from vendors. Using these inputs, \papername searches for the optimal architecture in three steps, as shown in Fig.~\ref{fig:searching}: 1) configure compute units and on-chip memory, 2) define inter-core or inter-device hardware specifications, and 3) evaluate performance using \papername scheduling methods. Each step is detailed below.

%%%%%%%%%%%%%%%%%%%%%%%%%%%%%%%%%%%%%%%%%%%%%%%%%%%%%%%%%%%%%%%%%%%%%%%%%%
%% Figure 7 : ADOR's Searching Method

%%%%%%%%%%%%%%%%%%%%%%%%%%%%%%%%%%%%%%%%%%%%%%%%%%%%%%%%%%%%%%%%%%%%%%%%%%

% Searching Methodology 
\subsection{Compute Units}

First, the ratio of compute units among the systolic array, MAC tree, and vector unit within the core is determined. The MAC tree is the first module to be allocated and must have enough compute units to process weights fetched from memory bandwidth each cycle. However, since there are operations like multi query attention~ (MQA)~\cite{almazrouei2023falcon} or group query attention (GQA)~\cite{jiang2023mistral} that reuse key-value pairs or MoE layers~\cite{fedus2022switch} that share the same experts, the initial allocation of MAC tree compute units is based on the formula provided below.

%%%%%%%%%%%%%%%%%%%%%%%%%%%%%%%%%%%%%%%%%%%%%%%%%%%%%%%%%%%%%%%%%%%%%%%%%%
%% Equation 1

\begin{figure}[h]
\centering
\vspace{-0.10in}
\includegraphics[width=1.0\linewidth]{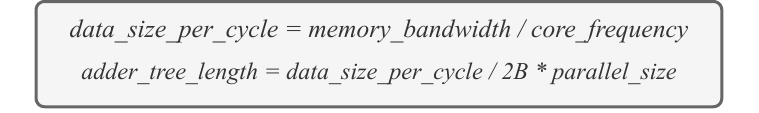}
\label{eq:adder-tree-size}
\vspace{-0.25in}
\end{figure}

%%%%%%%%%%%%%%%%%%%%%%%%%%%%%%%%%%%%%%%%%%%%%%%%%%%%%%%%%%%%%%%%%%%%%%%%%%

\noindent
If too many compute units are allocated to the MAC tree, there may not be enough left for the systolic array, which can negatively impact throughput.

%%%%%%%%%%%%%%%%%%%%%%%%%%%%%%%%%%%%%%%%%%%%%%%%%%%%%%%%%%%%%%%%%%%%%%%%%%
%% Figure 10 : Estimate Bandwidth Utilization for MAC tree

\begin{figure}[t]
\centering
\includegraphics[width=1.0\linewidth]{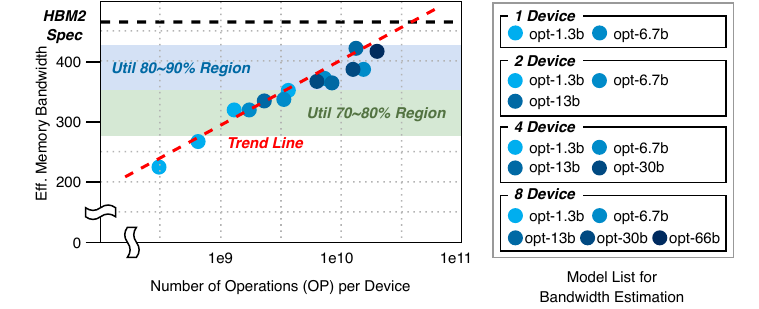}
\vspace{-0.25in}
\caption{Measurement of effective memory bandwidth for the MAC tree. The actual bandwidth was measured using a AMD U55C FPGA.}
\label{fig:band_util}
\vspace{-0.10in}
\end{figure}

%%%%%%%%%%%%%%%%%%%%%%%%%%%%%%%%%%%%%%%%%%%%%%%%%%%%%%%%%%%%%%%%%%%%%%%%%%
%% Figure 11 : Core Latency

\begin{figure}[t]
\centering
\includegraphics[width=1.0\linewidth]{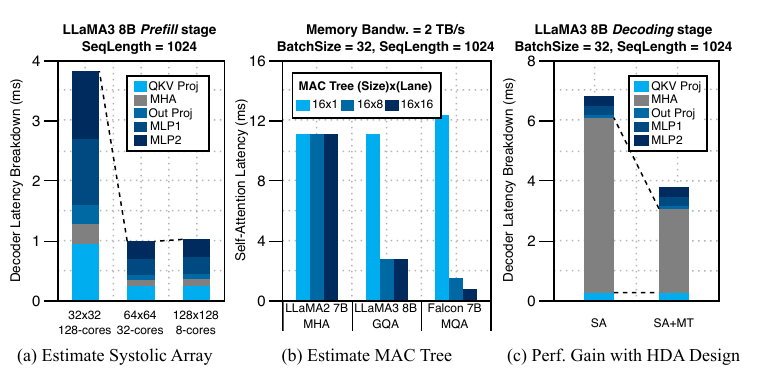}
\vspace{-0.25in}
\caption{(a) Performance comparison of various systolic array configurations (b) Performance comparison of MHA, GQA, and MQA based on the number of lanes in the MAC tree (c) Performance gains with HDA architecture}
\label{fig:core_latency}
\vspace{-0.20in}
\end{figure}

%%%%%%%%%%%%%%%%%%%%%%%%%%%%%%%%%%%%%%%%%%%%%%%%%%%%%%%%%%%%%%%%%%%%%%%%%%

To accurately predict the performance of \papername, it is crucial to determine the memory bandwidth utilization when using the MAC tree. However, measuring this through simulation alone can be extremely challenging and can result in significant performance estimation errors. To evaluate memory bandwidth utilization, we implemented the MAC tree on AMD’s Alveo U55C FPGA~\cite{alves2016xilinx}, which features two HBM2 modules with a total bandwidth of 460 GB/s. Fig.\ref{fig:band_util} demonstrates a logarithmic relationship between the computational workload of various LLM models and memory bandwidth utilization. This indicates that the memory bandwidth utilization can be accurately predicted based on the model’s computational workload. The MAC tree achieves up to 90\% of the theoretical maximum bandwidth, making it highly effective for latency-sensitive workloads. The number of lanes in the MAC tree can be determined by measuring the performance of various self-attention mechanisms, as shown in Fig.~\ref{fig:core_latency}-(b).

Once the MAC trees are determined, the remaining units are assigned to the systolic array. The configuration of the systolic array and the number of cores are then adjusted to find the optimal setup for efficiently processing GEMM operations. A too-small systolic array with many cores might not allow sufficient input reuse due to limited local SRAM size, while a too-large systolic array with few cores can lead to underutilized cores during tiling, reducing overall hardware utilization. Therefore, configurations are tested in multiples of 32 to find the best performing setup for GEMM operations.

To model the systolic array, we utilized SCALE-Sim~\cite{samajdar2018scale}. By analyzing the model structure and converting GEMM operations into a format readable by SCALE-Sim, we measured latency ratios for the given hardware configuration. Fig.~\ref{fig:core_latency}-(a) shows an example of performance measurements for various systolic array sizes using SCALE-Sim.

%%%%%%%%%%%%%%%%%%%%%%%%%%%%%%%%%%%%%%%%%%%%%%%%%%%%%%%%%%%%%%%%%%%%%%%%%%
%% Figure 12 : Local Memory Usage

\begin{figure}[t]
\centering
\includegraphics[width=1.0\linewidth]{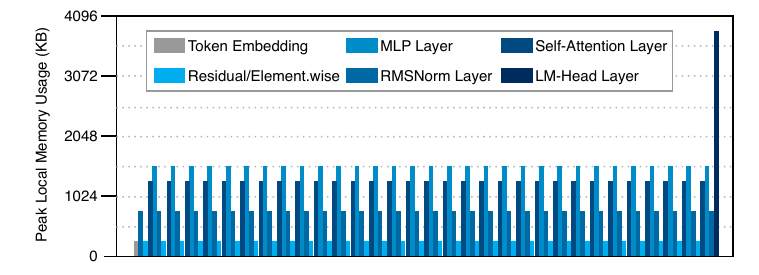}
\vspace{-0.25in}
\caption{Local memory usage for a batch size of 32 based on LLaMA3 8B. Except for the LM-Head, the usage does not exceed 1.5 MB.}
\label{fig:sram_usage}
\vspace{-0.20in}
\end{figure}

%%%%%%%%%%%%%%%%%%%%%%%%%%%%%%%%%%%%%%%%%%%%%%%%%%%%%%%%%%%%%%%%%%%%%%%%%%

\subsection{Local Memory \& Global Memory}

The size of the local memory is determined by the maximum memory usage of activations when executing the given model. Generally, the LM-Head and Self-Attention mechanisms require substantial local memory. The memory usage of the LM-Head is determined by the model’s vocab size, which is typically larger than the hidden size. However, since the LM-Head is only involved in the \textit{decoding} stage and not in the \textit{prefill} stage, its maximum usage can be predicted based on the batch size. For Self-Attention, the Score Matrix occupies a significant amount of memory, especially as the maximum sequence length has recently increased substantially. Fortunately, using memory optimization techniques like softmax decomposition from the FlashAttention~\cite{dao2022flashattention} library can help reduce memory usage. We have developed a simulator to calculate local memory usage, as shown in Fig.~\ref{fig:sram_usage}. For global memory, a larger capacity allows for storing more key-value pairs. Therefore, after determining the local memory size, the remaining SRAM is fully allocated to global memory.

%%%%%%%%%%%%%%%%%%%%%%%%%%%%%%%%%%%%%%%%%%%%%%%%%%%%%%%%%%%%%%%%%%%%%%%%%%
%% Figure ? : Scalabitliy

\begin{figure}[t]
\centering
\includegraphics[width=1.0\linewidth]{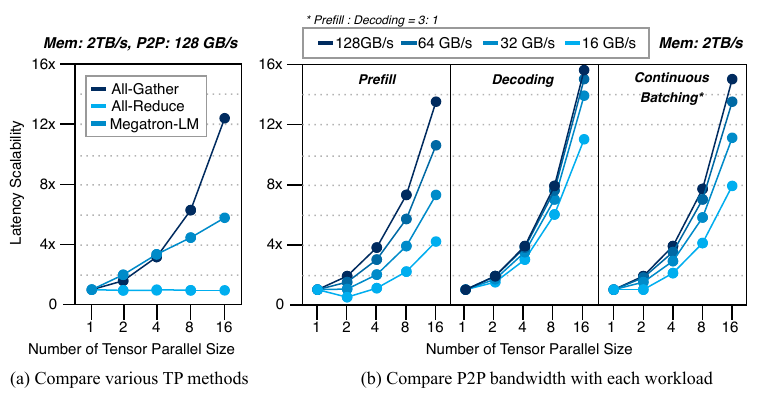}
\vspace{-0.25in}
\caption{(a) Scalability metrics for All-gather, All-reduce, and Megatron-LM methods. Megatron-LM performs best with fewer devices, but All-gather shows the highest scalability as the number of devices increases. (b) Scalability for \textit{prefill}, \textit{decoding}, and mixed workloads with various P2P bandwidths. For \textit{decoding}, due to memory-bound attention block, better overlapping tendencies are observed.}
\label{fig:scalability}
\vspace{-0.20in}
\end{figure}

%%%%%%%%%%%%%%%%%%%%%%%%%%%%%%%%%%%%%%%%%%%%%%%%%%%%%%%%%%%%%%%%%%%%%%%%%%

\subsection{NoC and P2P Specifications}
\label{sec:searching:noc_p2p}

Once the core architecture is determined, the required NoC and P2P specifications are established based on the dataflow. For latency-sensitive workloads, NoC bandwidth is set to minimize synchronization overhead by overlapping computation and communication during GEMV operations. For throughput-sensitive workloads, the bandwidth required to hide weight pre-fetching increases with the size of the systolic array. The final NoC bandwidth is the higher of these two values. If the required bandwidth is too high, based on recent studies~\cite{kwon2021heterogeneous, cai2024gemini}, adjustments are made to the number of cores and systolic array size in step 1).

To measure latency based on NoC bandwidth, we set up a tool capable of simulating scenarios where computation and communication overlap. The latency of computation can be calculated using the effective memory bandwidth obtained in the previous section, while communication latency can be measured based on the amount of data transmitted and the available bandwidth. This allows us to determine the minimum bandwidth required to ensure that computation and communication overlap effectively.

After determining the NoC specifications, the P2P bandwidth is set similarly. As discussed in Section~\ref{sec:multi_device}, the minimum bandwidth required for efficient communication and synchronization during parallelism is identified. We evaluate the scalability of each TP method, determining that Megatron is more efficient with two devices, while all-gather scales better with four or more devices (Fig.\ref{fig:scalability}). A bandwidth of approximately 32 GB/s, achievable with PCIe-4$\times$16, is sufficient for overlapping computation and communication. Unlike NVIDIA GPUs, which rely on high-bandwidth, high-power connections like NVLink\cite{li2019evaluating}, \papername enables efficient multi-device computing with PCIe or Infiniband~\cite{grun2010introduction}.

%%%%%%%%%%%%%%%%%%%%%%%%%%%%%%%%%%%%%%%%%%%%%%%%%%%%%%%%%%%%%%%%%%%%%%%%%%
%% Figure 8 : ADOR Simulator

\begin{figure}[t]
\centering
\includegraphics[width=1.0\linewidth]{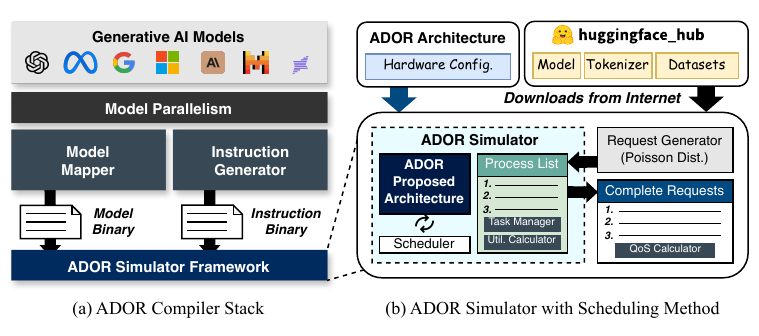}
\vspace{-0.25in}
\caption{(a) Complier stack of the \papername framework. (b) \papername simulator replicating a real LLM serving environment, capable of measuring QoS, hardware utilization, and maximum number of requests handled.}
\label{fig:simulator}
\vspace{-0.20in}
\end{figure}

%%%%%%%%%%%%%%%%%%%%%%%%%%%%%%%%%%%%%%%%%%%%%%%%%%%%%%%%%%%%%%%%%%%%%%%%%%

\subsection{Estimate Dynamic Scheduling}

Once the hardware configuration is finalized, the performance of \papername’s proposed hardware is evaluated in a simulated service environment. The \papername Simulator calculates metrics such as hardware utilization, throughput, and QoS indicators (TTFT, TBT, and E2E latency) to ensure both vendor and end-user requirements are met. Using the compiler stack shown in Fig.\ref{fig:simulator}-(a), instruction and memory-mapped files are generated for simulation. The simulator then downloads model information (e.g., decoder layers, attention heads) and service trend datasets from HuggingFace to reconstruct input and output token patterns (Fig.\ref{fig:simulator}-(b)). A \textit{Request Generator} simulates user requests with a Poisson distribution, and the scheduler assigns processes to hardware for execution, returning QoS metrics from completed requests.

If the simulation meets both vendor and end-user requirements, the architecture is finalized and proposed. Otherwise, the search restarts from step 1, allocating more resources to the systolic array for vendor needs or to the MAC tree for end-user needs. If requirements are still unmet after multiple iterations, the final architecture is proposed along with the additional hardware specifications needed.

%%%%%%%%%%%%%%%%%%%%%%%%%%%%%%%%%%%%%%%%%%%%%%%%%%%%%%%%%%%%%%%%%%%%%%%%%%
%%%%%%%%%%%%%%%%%%%%%%%%%%%%%%%%%%%%%%%%%%%%%%%%%%%%%%%%%%%%%%%%%%%%%%%%%%
%% Copyright HyperAccel, Inc. All Rights Reserved.
%% Section: Evaluation
%%%%%%%%%%%%%%%%%%%%%%%%%%%%%%%%%%%%%%%%%%%%%%%%%%%%%%%%%%%%%%%%%%%%%%%%%%

\section{Evaluation}
\label{sec:evaluation}

Through the \papername framework, we can propose hardware architectures that are efficient for serving LLM. In this section, we evaluate the performance of the hardware architectures proposed by \papername and compare them with existing hardware solutions to demonstrate their efficiency in providing LLM services.

%%%%%%%%%%%%%%%%%%%%%%%%%%%%%%%%%%%%%%%%%%%%%%%%%%%%%%%%%%%%%%%%%%%%%%%%%%
%% Table 2 : Configurations of ADOR Hardware

\begin{table}[t]
    \centering
    \caption{Hardware Specifications Proposed by \papername}
    \begin{tabular}{l|c|c|c|c}
        \toprule
        \multirow{2}{*}{\textbf{Specifications}} & NVIDIA & \multicolumn{2}{c}{LLMCompass~\cite{zhangllmcompass}} & \textbf{\papername} \\
        & A100 & L & T & \textbf{Design}\\
        \midrule
        Core Frequency (MHz) & 1500 & 1500 & 1500 & 1500 \\
        Systolic Array & - & 16$\times$16 & 32$\times$32 &  \textbf{64$\times$64} \\
        MAC Tree & - & - & - & \textbf{16$\times$16} \\
        Lane Count & - & 4 & 4 & 1 \\
        Core Count & 108 & 64 & 64 & \textbf{32} \\
        Local Memory (KB) & 192 & 192 & 768 & \textbf{2048} \\
        Global Memory (MB) & 48 & 24 & 48 & \textbf{16} \\
        DRAM Size (GB) & 80 & 80 & 512 & 80 \\
        DRAM Bandwidth (TB/s) & 2 & 2 & 1 & 2 \\
        P2P Bandwidth (GB/s) & 600 & 600 & 600 & \textbf{64} \\
        \hline
        Performance (TFLOPS) & 312 & 196 & 786 & \textbf{417} \\
        Die Area (7nm, mm$^2$) & 826 & 478 & 787 & \textbf{516} \\
        \bottomrule
    \end{tabular}
    \label{tab:compare_uarch}
    \vspace{-0.10in}
\end{table}

%%%%%%%%%%%%%%%%%%%%%%%%%%%%%%%%%%%%%%%%%%%%%%%%%%%%%%%%%%%%%%%%%%%%%%%%%%

% Experimental Setup
\subsection{Experimental Setup}
\label{sec:evaluation:setup}

For a fair comparison, \papername proposed hardware configurations with similar specifications as the A100. The performance was evaluated against the A100 as well as the state-of-the-art hardware architecture from the LLM inference simulator~\cite{zhangllmcompass}. \papername configured a MAC tree with a size of 16 based on a bandwidth of 2 TB/s, and set up 16 lanes to ensure sufficient parallelization. To closely match the number of compute units of the A100, a systolic array was configured to 64$\times$64. The number of cores and global memory size were adjusted to ensure adequate local memory capacity. The P2P bandwidth was configured to 64 GB/s, which is sufficient to avoid performance bottlenecks. To estimate the area of the proposed \papername architecture, we added the MAC tree information to the LLMCompass~\cite{zhangllmcompass} cost model.

%%%%%%%%%%%%%%%%%%%%%%%%%%%%%%%%%%%%%%%%%%%%%%%%%%%%%%%%%%%%%%%%%%%%%%%%%%
%% Figure ? : Evaluation

\begin{figure}[t]
\centering
\includegraphics[width=1.0\linewidth]{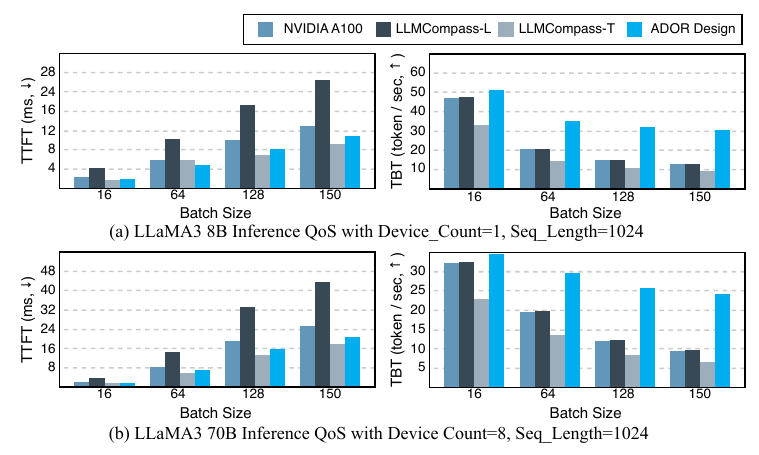}
\vspace{-0.25in}
\caption{Comparison of QoS between the design proposed by ADOR and other hardware. Measurements of TTFT and TBT for LLaMA3 8B and LLaMA3 70B with varying batch sizes.}
\label{fig:evalution}
\vspace{-0.20in}
\end{figure}

%%%%%%%%%%%%%%%%%%%%%%%%%%%%%%%%%%%%%%%%%%%%%%%%%%%%%%%%%%%%%%%%%%%%%%%%%%

\subsection{QoS Comparison with Other Hardware Designs}

Fig.~\ref{fig:evalution} compares the \papername design with the A100 and LLMCompass’s latency-oriented (LLMCompass-L) and throughput-oriented (LLMCompass-T) designs for LLaMA3 8B and 70B. For a batch size of 16, \papername performs similarly to the A100, but as the batch size increases, \papername maintains higher bandwidth utilization, outperforming the A100 in TBT. At a batch size of 150 for LLaMA3 8B, \papername achieves 2.36$\times$ higher TBT and 1.93$\times$ and 3.78$\times$ improvements in area efficiency for TTFT and TBT, respectively. For LLaMA3 70B using 8 devices, \papername shows 2.51$\times$ better TBT and 4.01$\times$ higher area efficiency. While LLMCompass excels in latency (LLMCompass-L) or throughput (LLMCompass-T), it lacks the balanced performance \papername delivers for both TTFT and TBT.

\subsection{QoS in Real-World LLM Serving}

\begin{figure}[t]
\centering
\includegraphics[width=1.0\linewidth]{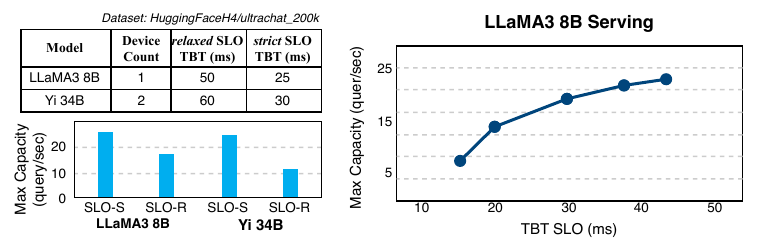}
\vspace{-0.25in}
\caption{Number of requests that can be maximally processed under a given SLO in the \papername serving environment. Measured in the chatbot service environment with LLaMA3 8B and Yi 34B models.}
\label{fig:slo}
\vspace{-0.20in}
\end{figure}

Fig.\ref{fig:slo} illustrates the performance of the proposed hardware design in a real LLM serving environment, where batch sizes dynamically change with user requests, and TTFT impacts latency alongside TBT due to overlapping \textit{prefill} and \textit{decoding} stages. Using the HuggingFaceH4/ultrachat\_200k~\cite{ding2023enhancing} dataset, we configured a chatbot service with LLaMA3 8B and Yi 34B models to measure the maximum requests \papername can handle under defined SLOs~\cite{agrawal2024taming}. Results show \papername achieves high throughput, with TTFT and TBT remaining stable as batch sizes increase, enabling rapid growth in maximum capacity.

\begin{figure}[t]
\centering
\includegraphics[width=1.0\linewidth]{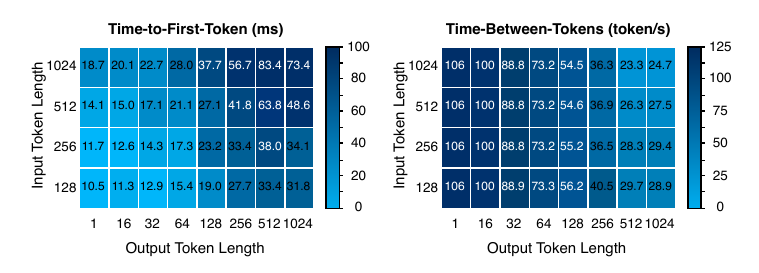}
\vspace{-0.25in}
\caption{QoS measurements for various sequence lengths when serving LLaMA3 8B with the \papername design.}
\label{fig:qos}
\vspace{-0.20in}
\end{figure}

Fig.~\ref{fig:qos} shows the QoS measurements for various sequence lengths when serving LLaMA3 8B with the \papername design. For TBT, as the token length increases, the overlap between \textit{prefill} and \textit{decoding} stages inevitably slows down the processing time per token. However, due to the MAC tree, the processing time decreases by only an average of 3.87$\times$ as the output token length increases from 1 to 1024. For TTFT, the \papername design also maintains high throughput and quickly handles the overlapping decoding stage, resulting in only an 3.85$\times$ decrease. This is 2.21$\times$ higher than the GPU under the same conditions, indicating that \papername better withstands QoS degradation with increasing sequence length.

%%%%%%%%%%%%%%%%%%%%%%%%%%%%%%%%%%%%%%%%%%%%%%%%%%%%%%%%%%%%%%%%%%%%%%%%%%
%% Figure ? : SLO

%%%%%%%%%%%%%%%%%%%%%%%%%%%%%%%%%%%%%%%%%%%%%%%%%%%%%%%%%%%%%%%%%%%%%%%%%%
%% Figure ? : QoS

%%%%%%%%%%%%%%%%%%%%%%%%%%%%%%%%%%%%%%%%%%%%%%%%%%%%%%%%%%%%%%%%%%%%%%%%%%

%%%%%%%%%%%%%%%%%%%%%%%%%%%%%%%%%%%%%%%%%%%%%%%%%%%%%%%%%%%%%%%%%%%%%%%%%%
%% Copyright HyperAccel, Inc. All Rights Reserved.
%% Section: Conclusion
%%%%%%%%%%%%%%%%%%%%%%%%%%%%%%%%%%%%%%%%%%%%%%%%%%%%%%%%%%%%%%%%%%%%%%%%%%

\section{Conclusion}
\label{sec:conclusion}

In this paper, we address the challenges of serving LLMs during the \textit{prefill} and \textit{decoding} stages, which demand significant computational power and memory bandwidth. We identify the inefficiencies of current hardware, such as GPUs, and propose \papername, a framework that balances throughput and latency using predefined templates for heterogeneous dataflow. In a real LLaMA3 8B serving environment, \papername achieves 23.3 requests per second while meeting SLOs, delivers 2.51$\times$ higher TBT than the A100 with large batches, and improves area efficiency by 4.01$\times$, offering a cost-effective, resource-efficient solution.

%%%%%%%%%%%%%%%%%%%%%%%%%%%%%%%%%%%%%%%%%%%%%%%%%%%%%%%%%%%%%%%%%%%%%%%%%%

\bibliographystyle{IEEEtranS}
\bibliography{refs}

\end{document}